\documentclass[amsmath,amssymb,twocolumn]{revtex4}
\usepackage{mathrsfs}
\usepackage{graphicx}
\usepackage{color}
\usepackage{cases}
\usepackage{array}

\begin{document}

\title{Critical region of topological trivial and nontrivial phases in interacting Kitaev chain with spatially varying potentials}

\author{Weijie Huang$^{1}$ and Yao Yao$^{1,2}$\footnote{Electronic address:~\url{yaoyao2016@scut.edu.cn}}}

\address{$^1$ Department of Physics, South China University of Technology, Guangzhou 510640, China\\
$^2$ State Key Laboratory of Luminescent Materials and Devices, South China University of Technology, Guangzhou 510640, China}

\date{\today}

\begin{abstract}
By using the variational matrix product state method, we numerically study the interacting Kitaev chain with spatially varying periodic and quasi-periodic potentials and the latter follows the Fibonacci sequence. The edge correlation functions of Majorana fermions and low-lying ground states are computed to explore the robustness of topological superconducting phase. It is found that the original topological nontrivial phase is separated into to two branches by an emergent topological trivial phase, as a result of the competition among spatially varying potential, electronic Coulomb interaction and chemical potential. The analysis of energy gap and occupation number together suggests that the spontaneous symmetry breaking and the lift of degeneracy in the topological trivial phase are enabled by a potential-induced Fracton mechanism, namely the pairing of four Majorana fermions. It can be broken by further enhancing the interaction, and then the nontrivial phase reemerges. The evolution from the emergent fractal structure of population outside the critical region to the original structure of charge density wave is investigated as well.
\end{abstract}

\maketitle

\section{Introduction}

In the last decade, Majorana zero mode (MZM), a special kind of fermion, has attracted much attention in condensed matter physics, quantum computations, and other related fields \cite{beenakker2013search,elliott2015colloquium,leijnse2012introduction,wilczek2009majorana}. MZMs obey the non-abelian statistics and have promising potential to act as robust qubits for quantum computations \cite{alicea2012new,nayak2008non}. A number of realistic systems have been proposed to probably host MZMs \cite{kitaev2001unpaired,sarma2006proposal,moore1991nonabelions,fu2008superconducting,oreg2010helical,lutchyn2010majorana,sau2010generic,tewari2010theorem,alicea2010majorana,choy2011majorana,nadj2013proposal}
and extensive experimental efforts have been made in these systems \cite{mourik2012signatures,rokhinson2012fractional,das2012zero,deng2012anomalous,churchill2013superconductor,lee2014spin,nadj2014observation,wang2012coexistence,xu2015experimental,sun2016majorana}.
On the theoretical side, Kitaev introduced a celebrated model based upon the spinless p-wave superconductor in one dimension \cite{kitaev2001unpaired}, which is currently named as the Kitaev model, one of the simplest but highly nontrivial model to unveil MZM. The most appealing physics that the Kitaev model tells us is the Majorana edge modes which are exponentially localized at the ends of the chain. By these edge modes, nonlocal fermionic excitations with zero energy emerge and the ground state becomes twofold degenerate. The quantum computations based upon these zero-energy excitations are then regarded to be undissipative and robust.

Driven by the demand of applications, researchers devoted great effort to the stability of the topological phase in the Kitaev chain with nearest-neighboring repulsive interactions \cite{wu2014new,amorim2015majorana,alicea2011non}. Along this line, theoretical studies also extended to issues of dimerization \cite{wakatsuki2014fermion}, disorder \cite{PhysRevB.90.121407,PhysRevLett.107.196804,PhysRevLett.109.146403,PhysRevLett.110.146404,PhysRevLett.112.206602,PhysRevB.88.165111}, quasi-periodicity \cite{PhysRevB.85.140508,PhysRevLett.110.146404}, long-range interactions \cite{PhysRevB.93.165423,ohta2015phase,mahyaeh2018zero}, quartic interactions \cite{stoudenmire2011interaction}, and so on. For the interacting model without disorders \cite{PhysRevB.84.085114,hassler2012strongly,PhysRevB.88.161103,PhysRevB.92.115137}, both numerical and perturbative investigations showed that MZMs can be stably present with moderate interactions \cite{PhysRevB.84.014503,PhysRevB.84.085114,hassler2012strongly,PhysRevB.88.161103,manolescu2014coulomb,PhysRevB.91.165402}, which was also found to generally broaden the window of chemical potential where the system is in the nontrivial topological superconducting (TSC) phase \cite{PhysRevB.84.014503,PhysRevB.84.085114,hassler2012strongly,PhysRevB.88.161103,manolescu2014coulomb}. In addition, MZMs were found to survive broader parameter regions in disordered/quasi-periodic chains as well \cite{PhysRevB.63.224204,PhysRevLett.110.146404,PhysRevLett.110.176403}. It is thus interesting what would take place in presence of both interactions and disorders. This is meaningful also in the applicable manner if experimenters want to freely manipulate Majorana modes in one-dimensional wires \cite{PhysRevB.90.121407,PhysRevB.93.075129,PhysRevLett.109.146403,PhysRevB.86.205135,lobos2012interplay,crepin2014nonperturbative}. So far, it has been proven that moderate disorders and repulsive interactions together are able to stabilize the topological order, but when the disorders are sufficiently strong, both repulsive and attractive interactions suppress the topological phase \cite{PhysRevB.93.075129}.

In realistic materials, moderate disorders can be realized as quasi-periodic spatially varying potentials. The simplest model for quasi-periodicity is the Harper model \cite{azbel1964energy} with a cosine-like shaped potential, which has been considered in Kitaev chains \cite{PhysRevLett.110.146404,PhysRevLett.110.176403,PhysRevB.86.205135,PhysRevB.85.140508}. It can also be demonstrated that this model is closely related to the anisotropic XY spin chain (AXYSC) through the Jordan-Wigner (J-W) transformation \cite{wada2021coexistence,PhysRevB.101.125431,PhysRevB.102.165147,PhysRevB.101.045422,fendley2016strong,PhysRevB.100.235127,saha2019characterization,PhysRevB.100.104428}. One is then interesting in more nontrivial case. So a potential with Fibonacci sequence is introduced to the Kitaev chain which is the so-called Fibonacci-Kitaev chain \cite{ghadimi2017majorana}. On the basis of this model, there is a great chance to generate new self-similar fractal structures with regard to Majorana zero-energy mode. On the other hand, the Kitaev chains with density-density interaction (Kitaev-Hubbard chain) and the quantum axial next-nearest-neighbor Ising (ANNNI) model \cite{PhysRevB.73.052402,PhysRevB.76.094410,nagy2011exploring,PhysRevB.84.014407,allen2001two} are regarded to be dual to each other, and a new topological phase was found while using bosonization and DMRG to investigate the phase diagram of the model \cite{PhysRevB.101.085125}. Among all these researches, various fingerprints have been employed to quantify the phase diagram, including entanglement entropy and entanglement spectrum \cite{ohta2016topological,PhysRevB.93.075129,PhysRevB.101.085125,PhysRevB.96.241113,stoudenmire2011interaction,PhysRevB.92.235123,PhysRevB.101.125431}, Lyapunov exponent \cite{ghadimi2017majorana,PhysRevLett.110.146404,monthus2018topological}, string correlation function \cite{ohta2016topological,PhysRevB.93.075129,miao2018majorana,PhysRevLett.110.176403,PhysRevLett.118.267701,PhysRevB.92.115137}, many-body Majorana operator \cite{PhysRevB.92.081401,goldstein2012exact}, Hartree-Fock analysis \cite{PhysRevB.84.014503}, lowing-energy spectrum/gap \cite{ohta2016topological,degottardi2011topological,PhysRevB.101.085125,wada2021coexistence,PhysRevB.100.235127,PhysRevLett.107.196804,PhysRevB.92.235123,PhysRevB.95.195140,PhysRevB.98.155119}, and tunneling spectra \cite{PhysRevB.88.161103}.

In this paper, we study the interacting Kitaev chain with spatially varying potentials. We assign the chemical potential at certain sites to be zero and leave the others to have a certain value. The sites with zero chemical potential follow the periodic or quasi-periodic sequence. By using the variational matrix product state (VMPS) method, also known as the matrix product states version of density-matrix renormalization group (DMRG), we calculate the edge correlation function of the two Majorana operators at the edges and use it as a long-range order parameter to characterize the nontrivial TSC phase with MZMs. We find the symmetry is spontaneously broken and a topological trivial phase appears to divide the TSC phase into two branches. By calculating the low-lying ground state of two different parity sectors, we find this topological trivial phase has a non-degenerate ground state \cite{miao2018majorana}. We also calculate the occupation with different parameters. Results are in good agreement in the two cases: the potential of periodic sequence and quasi-periodic Fibonacci sequence. These chains have adjacent sites with zero chemical potential, so it is closely relevant to the Fracton physics \cite{PhysRevB.102.214437,nandkishore1803fractons,PhysRevResearch.1.013011,PhysRevResearch.1.013011,PhysRevB.92.235136,PhysRevB.103.085101}.

\section{MODEL}

\subsection{Kitaev model and J-W transformtion}

Let us begin with the benchmarking Kitaev model for spinless fermions with open boundary condition \cite{kitaev2001unpaired}. The Hamiltonian is written as
\begin{eqnarray}
H&=&\sum_{j=1}^{L-1}\left[-t\left(c_{j}^{\dagger} c_{j+1}+\rm{h.c.}\right)+U\left(2 n_{j}-1\right)\left(2n_{j+1}-1\right)\right.\nonumber\\
&-&\left.\Delta\left(c_{j}^{\dagger} c_{j+1}^{\dagger}+\rm{h.c.}\right)\right]- \sum_{j=1}^{L}\mu_j\left(n_{j}-\frac{1}{2}\right),\label{eq:one}
\end{eqnarray}
where the operator $c^\dagger_{j}$ $(c_{j})$ creates (annihilates) a spinless fermion on $j$-th site, $n_{j} = c^{\dagger}_{j}c_{j}$ is the corresponding fermion occupation operator,
$t$ is the hopping amplitude, $\Delta$ is the p-wave superconducting pairing potential, $\mu_j$ is the chemical potential on $j$-th site, and $U$ is the nearest-neighbor interaction. Without loss of generality, we can assume that $t$ and $\mu$ to be real and positive, since $t\rightarrow-t$ and $\mu\rightarrow-\mu$ and be realized by the gauge transformation $c_{j}\rightarrow i(-1)^{j}c_{j}$ and particle-hole conjugation $c_{j}\rightarrow (-1)^{j}c^\dagger_{j}$, respectively, and these transformations do not change other parameters.

It is well known this Hamiltonian can be represented in the Majorana fermion form. That is, a complex fermion operator can be split into two Majorana fermions operators:
\begin{eqnarray}
c_{j}=\frac{1}{2}\left(\lambda_{j}^{1}+i \lambda_{j}^{2}\right),
\end{eqnarray}
\begin{eqnarray}
c^{\dagger}_{j}=\frac{1}{2}\left(\lambda_{j}^{1}-i \lambda_{j}^{2}\right).
\end{eqnarray}
The Majorana fermion operators satisfy the Majorana condition $\left(\lambda_{j}^{a}\right)^{\dagger}=\lambda_{j}^{a}$ and also obey the anticommutation relation $\left\{\lambda_{j}^{a}, \lambda_{l}^{b}\right\}=2 \delta_{a b} \delta_{j l}$, where $a,b = 1,2$. So the Hamiltonian~(\ref{eq:one}) can be transformed to the following form:
\begin{eqnarray}
H&=&\sum_{j = 1}^{L-1}\left[-\frac{i}{2}(t+\Delta) \lambda_{j+1}^{1}
\lambda_{j}^{2}-\frac{i}{2}(t-\Delta) \lambda_{j}^{1}
\lambda_{j+1}^{2}
\right.
\nonumber\\
&-&\left.U \lambda_{j}^{1} \lambda_{j}^{2} \lambda_{j+1}^{1}
\lambda_{j+1}^{2}\right]-\frac{i}{2}  \sum_{j = 1}^{L}
\mu_{j}\lambda_{j}^{1} \lambda_{j}^{2}.
\label{eq:two}
\end{eqnarray}
Furthermore, one can use the J-W transformation to construct spin operators:
\begin{eqnarray}
S_{j}^{x} & = & \frac{1}{2} \lambda_{j}^{1} e^{i \pi \sum_{l<j} n_{l}}, \\
S_{j}^{y} & = & -\frac{1}{2} \lambda_{j}^{2} e^{i \pi \sum_{l<j} n_{l}} ,\\
S_{j}^{z} & = & \frac{i}{2} \lambda_{j}^{1} \lambda_{j}^{2}.
\end{eqnarray}
When $\Delta = t$, the Hamiltonian~(\ref{eq:two}) can be further mapped to a spin-chain Hamiltonian which is written in terms of spin operators $S_{j}^{x}$ and $S_{j}^{z}$, i.e.,
\begin{eqnarray}
H =  \sum_{j=1}^{L-1}-4 t S_{j}^{x} S_{j+1}^{x}+4 U S_{j}^{z} S_{j+1}^{z}
+\sum_{j=1}^{L} \mu_{j} S_{j}^{z}.
\end{eqnarray}
This form of Hamiltonian is friendly to the numerical approaches.

\subsection{Symmetries}

In presence of the pairing term, the total fermion number $\hat{N}=\sum_{j}n_{j}$ is not conserved. However, the Hamiltonian commutes with the fermion number parity $Z^{f}_{2}$ defined as
\begin{eqnarray}
Z^{f}_{2} = e^{i\pi\sum_{j} n_{j}} = (-1)^{\hat{N}}.
\end{eqnarray}
In addition, the particle-hole symmetry can be characterized by the particle-hole conjugation operator defined as
\begin{eqnarray}
Z_{2}^{p}=\prod_{j}\left[c_{j}+(-1)^{j} c_{j}^{\dagger}\right].
\end{eqnarray}
which is also conserved if and only if $\mu=0$. We can use these two symmetries $Z^{f}_{2}$ and $Z^{p}_{2}$ of the ground state to distinguish different phases.

\subsection{Quasi-periodic potential}

In the present work, we mainly focus on the lattices with chemical potential following the quasi-periodic Fibonacci sequence \cite{PhysRevB.51.6096}. One can use the following recursion formula to get the sequence composed of two symbols A and B. That is, we use the recursion formula $J_{n+1} = \{J_{n},J_{n-1}\}$, $n\ge 1$, $J_{0}= \{A\}$, $J_{1} = \{B\}$, so we have $J_{2} = \{J_{1},J_{0}\} = \{B,A\} $, $J_{3}=\{J_{2},J_{1}\} = \{B,A,B\} $,\dots. The total number of symbols in $S_{n}$ is given by the Fibonacci numbers $F_{n+1} = F_{n} + F_{n-1}$.

The models under study in the next Section are the interacting Kitaev chains with different types of spatial varying chemical potentials. We assign each $\mu_{i}$ by either $\mu_{A}$ or $\mu_{B}$. The periodicity and quasi-periodicity in $\mu_{i}$ are given by making the order of A and B follow the periodic or quasi-periodic sequences. For the actual value of $\mu_{A}$ and $\mu_{B}$ in the following calculation, we will set one of them to be zero ($\mu_{A}= 0$ or $\mu_{B}= 0$) and vary the other one. The chemical potential $\mu_{i}$ at adjacent sites might both be zero in some situations. This will presumably lead to the novel results.

\section{Results}

\begin{figure}
\includegraphics{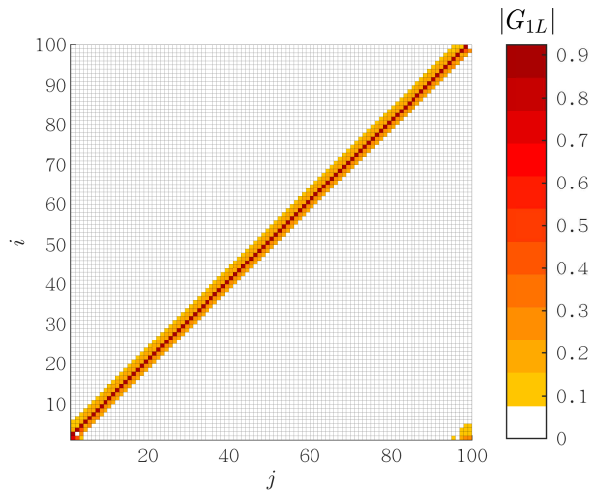}
\caption{\label{fig:G_ij} Correlation function $G_{ij}$ for the TSC ground state with $\Delta =t$, $\mu = 0$, $U=0.5t$, $L = 100$.}
\end{figure}

We use variational matrix product state (VMPS) \cite{schollwock2011density,itensor} to study the interplay of the nearest interaction and quasi-periodic chemical potential. We compute several observables, including the energy of low-lying states, local particle density, and most importantly the edge correlation function \cite{miao2018majorana}. The correlation function between two sites $i$ and $j$ is defined as
\begin{eqnarray}
G_{i j} & = & \left\langle i \lambda_{i}^{j} \lambda_{l}^{2}\right\rangle.
\end{eqnarray}
In particular, when $i = 1$ and $j = L$ it is the edge component of $G_{ij}$, i.e.,
\begin{eqnarray}
G_{1 L} & = & \left\langle i \lambda_{1}^{1} \lambda_{L}^{2}\right\rangle,
\end{eqnarray}
which is straightforwardly related to the edge modes. A typical result of $G_{ij}$ is shown in Fig.~\ref{fig:G_ij}. It is worth noting that the correlation function $G_{ij}$ is a block matrix of electron or hole density, which can be generalized to interacting systems and reflects the site-distribution of single-particle elementary excitations in a many-body ground state. As long as the bulk is homogeneous, in the thermodynamic limit a finite value of $G_{1L}$ fingerprints the existence of edge modes, since the correlation can not be transferred site by site to such long distance. One may then wonder how about the inhomogeneous lattice? In this work, we thus calculate $G_{1L}$ with spatially varying chemical potential, which will lead the bulk to be inhomogeneous. The nonvanishing edge correlation function $G_{1L}=\left\langle i \lambda_{1}^{1} \lambda_{L}^{2}\right\rangle$ characterizes the topological order, that is, the value $G_{1L}$ is finite in TSC phase and vanishes in other topological trivial phases, and also this order parameter is valid both in non-interacting and interacting systems. Fixing $\Delta = t$, we can plot the ground state edge correlation $G_{1L}$ of the interacting Kitaev chains as a function of $U$.

\begin{figure}
\includegraphics{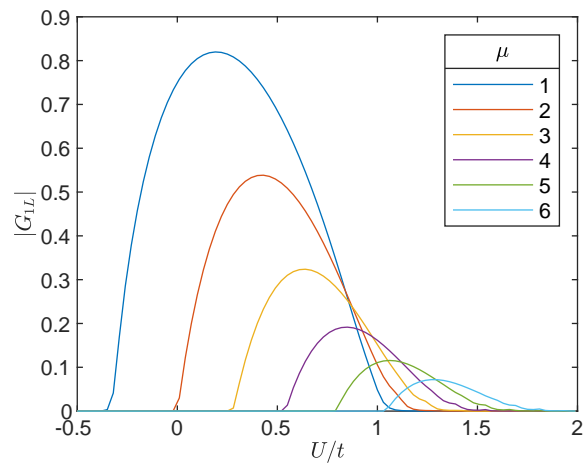}
\caption{\label{fig:G1L_88_normal_v2} Edge correlation function $G_{ij}$ of the ground state of the interacting Kitaev chain as a function of $\mu$ and $U$ with $\Delta = t$ and $ L = 88$.}
\end{figure}

\subsection{Periodic chemical potential}

\begin{figure}
\includegraphics{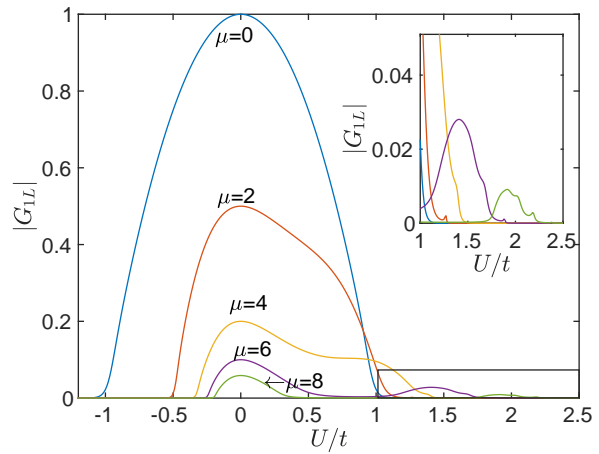}
\caption{\label{fig:G1L_88_Periodic}
Edge correlation function $G_{ij}$ of the ground state with periodic chemical potential as functions of $\mu$ and $U$. $L = 88$, $\Delta = t$, [$\mu$ , 0, 0] as a period repeat form site 1 to 87, and $\mu_L = \mu$. Region in the black box is zoomed in and illustrated in the insert. }
\end{figure}

We first calculate the edge correlation function of the interacting Kitaev chain with 88 sites. Two cases without and with periodic chemical potentials are considered, with the results shown in Fig.~\ref{fig:G1L_88_normal_v2} and Fig.~\ref{fig:G1L_88_Periodic} respectively for comparison. We notice that, since the model hosts MZMs on the edges, the chemical potential on the end site of the chain is extremely important \cite{PhysRevB.101.085125}. The chemical potential is repeated by [$\mu$, 0, 0] from site 1 to 87 and the final site is set to $\mu_{L} = \mu$, the same as that on the first site. The results shown in the two figures are remarkably different. In Fig.~\ref{fig:G1L_88_normal_v2}, increasing the chemical potential $\mu$ makes the maximum value of $G_{1L}$ decrease and shifts the regions with nonvanishing $G_{1L}$ to the right where the interaction is stronger. In Fig.~\ref{fig:G1L_88_Periodic}, however, $G_{1L}$ manifests its maximum value with no interaction and survives even large chemical potential. More importantly, for a given $\mu$ there is a valley between two peaks, indicating the TSC phase is separated into two branches. This novel phenomenon means the phase diagram changes to a new pattern which is rather different from that in previous researches \cite{miao2018majorana,PhysRevB.88.161103,PhysRevB.92.115137,PhysRevB.101.085125,hassler2012strongly}. It is noteworthy that this effect is more evident with larger chemical potential.

\begin{figure}
\includegraphics{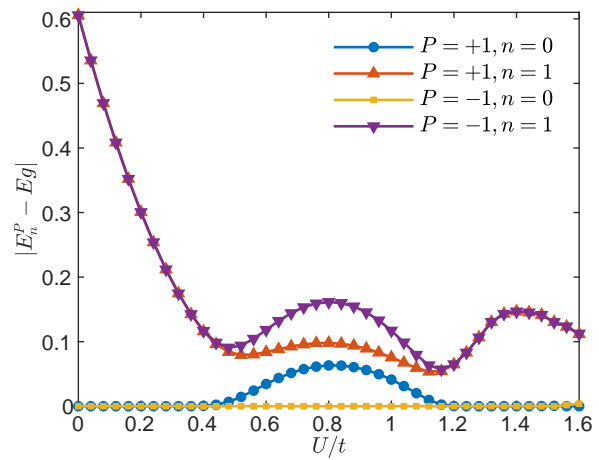}
\caption{\label{fig:EnergyGapPeriodic} Energy with respect to the ground state $E^{p}_{n}-E_{g}$. The ground state $(n=0)$ and the first excited state $(n=1)$ of two parity sectors $(Z^{p}_{2} = P = +1, -1)$ are calculated with $L = 88$ and $\mu = 6$.}
\end{figure}

As the VMPS or DMRG method often encounter the boundary problem that will affect the accuracy of using the edge correlation function to determine the phase transition point. We then analyze the symmetries in the ground state phase transition for comparison. The fermion number parity $Z_{2}^{f}$ conserves in this model which means the Hilbert space will be divided into two parity sectors denoted by $Z_{2}^{f} = +1, -1$. The TSC phase has opposite fermion number parity $Z_{2}^{f}$ in the twofold degenerate ground states while the charge density wave (CDW) and incommensurate charge density wave (ICDW) phase have the same $Z_{2}^{f}$. The lowest two eigenstates, which are the ground state ($n = 0$) and the first excited state ($n = 1$) in both parity sectors (labeled by $Z_{2}^{f} = P = +1, -1$) is represented in Fig.~\ref{fig:EnergyGapPeriodic}. The analysis of symmetries of the ground state consists with the results of the edge correlation function that the TSC phase does split into two. It means a spontaneous symmetry breaking takes place, and no degeneracy is found between the two TSC phases which is completely different from all the phases in the uniform interacting Kitaev chain \cite{miao2018majorana}. We will discuss these exotic phenomena extensively below.


\subsection{Quasi-periodic chemical potential}

\begin{figure}
\includegraphics{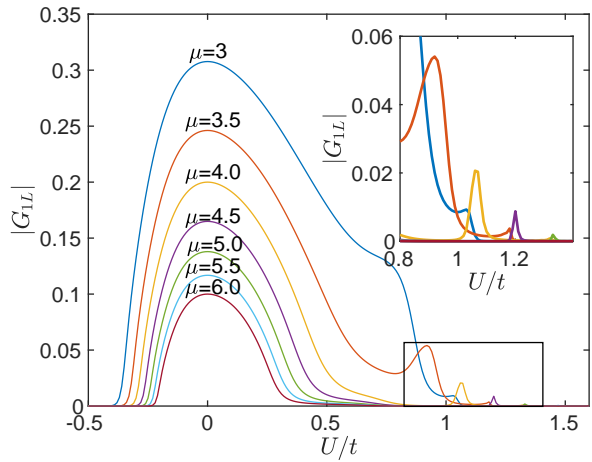}
\caption{\label{fig:Fibonacci_N=88} Edge correlation function $G_{ij}$ of the ground state of Fibonacci-Kitaev chain as functions of $\mu$ and $U$ with $ L = 88$, $\Delta = t$, $\mu_A=\mu$, $\mu_B = 0$, and $\mu_1 = \mu_L =\mu$. Region in the black box is zoomed in and illustrated in the insert.}
\end{figure}

We are now interesting in the situations if we use quasi-periodic sequences to substitute the periodic sequence. As it equivalently introduces moderate disorders to the system, one may image the split of the TSC phase probably would not show up. To this end, we use the Fibonacci sequence $J_{10}$ to generate quasi-periodic chemical potential and discard the first symbols in $J_{10}$, so the chain contains 88 sites and has non-zero chemical potential on both ends. The results are shown in Fig.~\ref{fig:Fibonacci_N=88}. It is found that, all the results are similar with that in the periodic case displayed in Fig.~\ref{fig:G1L_88_Periodic}. The edge function $G_{1L}$ with $\mu > 3$ also behaves a second growth as the repulsive interaction increases, where $G_{1L}$ can even decrease to around zero between two peaks for several $\mu$ values. Fixing $\mu$, there is a trivial region between two topological regions that is completely identical to the periodic case. Nonetheless, one can observe it more clearly in Fig.~\ref{fig:Fibonacci_N=88} that the TSC phase gradually breaks into two branches with the chemical potential increasing. The quasi-periodicity does not essentially kill or suppress the spontaneous symmetry breaking and the emergency of two topological regions. The TSC becomes even stabler than that in the uniform chain in small interaction region. The moderate disorder brought by the quasi-periodic potential broadens the chemical potential window in the non-interacting chain ($U = 0$), same with the previous results \cite{PhysRevB.84.014503,PhysRevB.84.085114,hassler2012strongly,PhysRevB.88.161103,manolescu2014coulomb}. Chains with more sites are calculated as well. That is, we use $J_{11}$ and $J_{12}$ to generate Fibonacci chains with 144 and 232 sites. The $G_{1L}$ around the phase boundary will be smaller making it more explicit for us to distinguish the split. We do not show results for more sites since the numerical calculation is hard to converge even if we kept bond dimension $\kappa = 400$ and run 100 sweeps.

\begin{figure}
\includegraphics{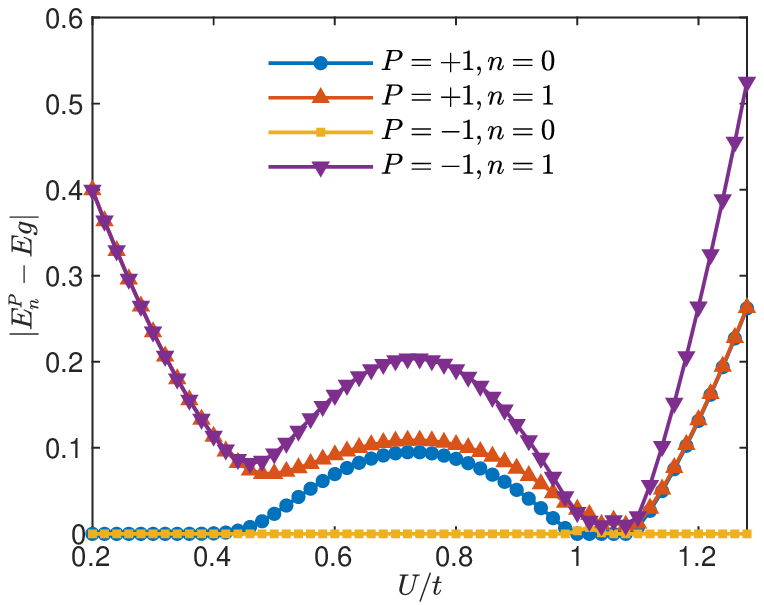}
\caption{\label{fig:Energy_Gap} Energy with respect to the ground state $E^{p}_{n}-E_{g}$. The ground state $(n=0)$ and the first excited state $(n=1)$ of two parity sectors $(Z_{2}^{p} = P = +1, -1)$ are ploted as a function of $U$, $L = 88$ and $\mu = 4$.}
\end{figure}

The lowest two eigenstates of the Fibonacci-Kitaev chain in both parity sectors are displayed in Fig.~\ref{fig:Energy_Gap}. Unlike the edge correlation function, the eigenstates for the quasi-periodic case is fairly different with the periodic case. For $U \gtrsim 0.6$ and $0.8 \lesssim U \lesssim 1 $, the ground state is doubly degenerate with opposite parity implying the chain is in a gapped topological phase. For $0.6 \lesssim U \lesssim 0.8 $, the degeneracy is lifted in this critical region and some symmetry is broken which might prefer a new topological trivial phase without superconductivity. The appearance of this exotic phase leads to the split of the TSC phase. The energy gap of this symmetry-breaking topological trivial phase is incredibly small, which is of the order $10^{-4}$ per site in the unit of hopping amplitude $t$. For $ U \gtrsim 1$ the $Z_{2}^{f} = 1$ sector has degenerate ground states but the other parity sector has got a gap between the ground state and the first excited state. In other words, the ground state is unique in this critical region, but will eventually have doubly degenerate states with different party if we keep increasing the interaction $U$.

We notice that there has been a research focusing on the region near the phase boundary and discussing the degeneracy and the parity of the ground state for different system sizes \cite{PhysRevB.101.085125}. Apart from the previously reported phases, a newly-observed ``excited state charge density wave" (esCDW) phase is present in that work. A more interesting point is that the esCDW phase appears only for even system sizes and is sensitive to the chemical potential at the edges. Furthermore, the transition point from this new phase to the CDW phase can actually governed by the chemical potentials at the two edges. The unique ground state in Fig.~\ref{fig:Energy_Gap} is found as well. More analysis of two lowest states in each parity sector can be found in this research. An exotic non-degenerate phase without the limitation of system size is found when we use the nonuniform chemical potentials. Our results turn out to be an exotic finding in addition to it.

\subsection{Critical region}

\begin{figure}
\includegraphics{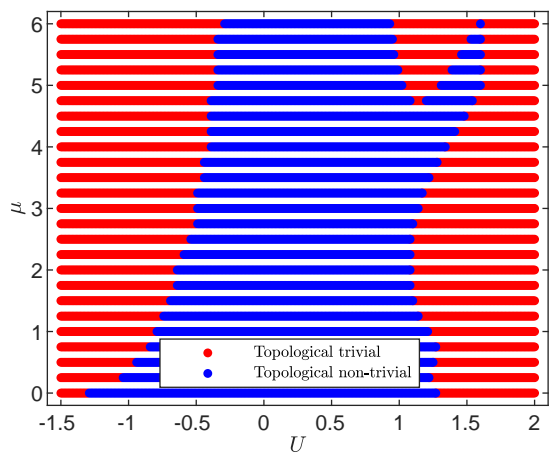}
\caption{\label{fig:TopologicalRegion} Phase diagram for the Fibonacci-Kitaev chain. Data points obtained with $ L = 88$, $\Delta = t$, $\mu_A = \mu$, $\mu_B = 0$, and $\mu_1 = \mu_L = \mu$. The edge correlation function $G_{ij}$ of the blue dots is nonvanishing which is in the topological superconductor phase and $G_{ij}$ of the red dots are less than $10^{-6}$ that is in topological trivial phase.}
\end{figure}

The stability of the topological order is related to all terms in the model, and the exact phase boundary is pretty difficult to determine. The phase diagrams of the uniform chain obtained by various methods have essentially the same pattern \cite{PhysRevB.84.014503,PhysRevB.84.085114,hassler2012strongly,PhysRevB.88.161103,manolescu2014coulomb}. Using the edge correlation function as a criterion, different phases can be then quantitatively assigned. The resulting phase diagram of chains with Fibonacci quasi-periodic potential is subsequently sketched in Fig.~\ref{fig:TopologicalRegion}. It has several fascinating differences from that for the uniform chain. As displayed in Fig.~\ref{fig:G1L_88_normal_v2}, the TSC phase of the uniform interacting Kitaev chain with larger $\mu$ tends to demand stronger interaction, where the $\mu = 0$ curve has got slight overlap with the $\mu = 6$ curve. But in Fig.~\ref{fig:TopologicalRegion} the TSC phase with different $\mu$ is restricted within a small range in the horizontal axis and piled up in the vertical axis. Especially, it is obvious in former results that there is an upper limit of chemical potential for the TSC phase to survive when the interaction strength is zero ($U = 0$). However, this limit disappears in Fig.~\ref{fig:TopologicalRegion} and the system can still stay in the TSC phase with Fibonacci quasi-periodic chemical potential up to an extremely large value. And also we can find for $\mu > 4.75$, the topological phase is divided into two branches with a topological trial phase in between. That is to say the second peak in Fig.~\ref{fig:Fibonacci_N=88} will result in the peninsula-like area in the phase diagram, which is the most distinctive difference from the uniform chain \cite{PhysRevB.84.014503,PhysRevB.84.085114,hassler2012strongly,PhysRevB.88.161103,manolescu2014coulomb}.

\begin{figure}
\includegraphics{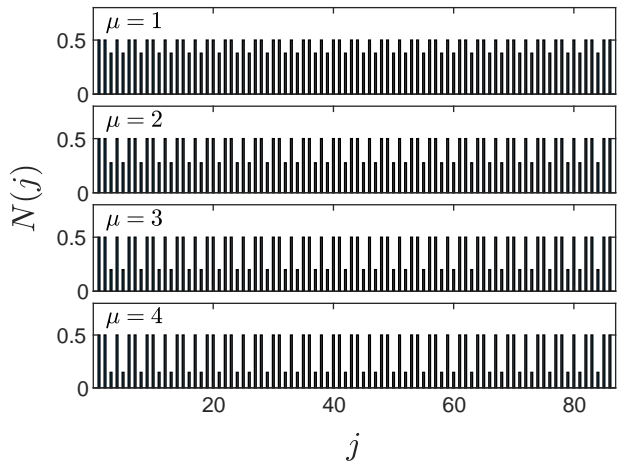}
\caption{\label{fig:Occupation} Occupation number $N(j)=\left \langle n_{j}\right \rangle$ of Fibonacci-Kitaev chain without interaction $(U = 0)$ of $\mu = 1, 2, 3, 4$. The system size $ L = 88$, $\Delta = t$, $\mu_a=\mu$, $\mu_b = 0$, and $\mu_1 = \mu_L = \mu$.}
\end{figure}

\begin{figure}
\includegraphics{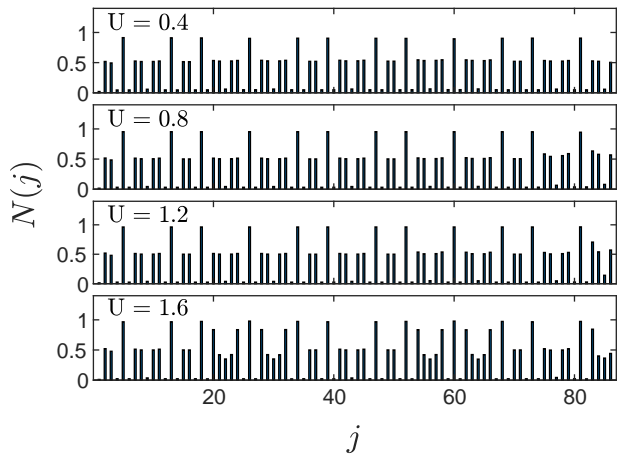}
\caption{\label{fig:OccupationMu=6} Occupation number $N(j)=\left \langle n_{j}\right \rangle$ of interacting Fibonacci-Kitaev chain with $\mu = 6$ and interaction $U = 0.4, 0.8, 1.2, 1.6$. The system size $ L = 88$, $\Delta = t$,
$\mu_a=\mu$, $\mu_b = 0$, and $\mu_1 = \mu_L = \mu$.}
\end{figure}

In the uniform interacting Kitaev chain, the occupation number decays drastically following $\mu$ increasing. The repulsive interaction prefers the ground states with occupation number having patterns like $(1010\cdots)$ and $(0101\cdots)$, which is expected to find in the phase of CDW. As the interaction becomes stronger the total occupation number will approach $N_{tot} = L/2$, the chemical potential will compete with the repulsive interaction. From Fig.~\ref{fig:Occupation} one can see the sites with $\mu_b = 0$ are half-occupied. We observe that the adjacent sites with $\mu = 0$ collect four Majorana fermions as a group, allowing the two fermions to move along the chain just like the mechanism of Fractons \cite{PhysRevB.102.214437,nandkishore1803fractons,PhysRevResearch.1.013011,PhysRevResearch.1.013011,PhysRevB.92.235136,PhysRevB.103.085101}. Consequently, the total occupation number is larger than in the normal interacting chain that the chemical potential window is broadened even further. It is intriguing that one can find the occupation number is nearly the same for $U = 0.4, 0.8, 1.2$ in Fig.~\ref{fig:OccupationMu=6}, namely, the topological trivial phase with $U = 1.2$ has the same occupation number distribution as the two branches of the nontrivial phase adjacent to it. More thorough investigations are needed for us to identify and understand properties of the exotic topological trivial phase.

\begin{figure}
\includegraphics{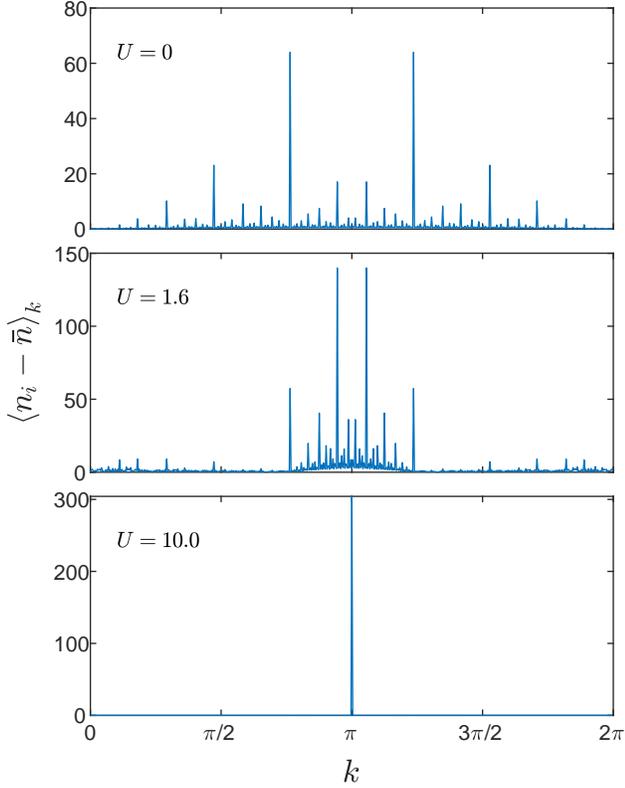}
\caption{\label{fig:Fourier} Fourier spectrum of the occupation number $N(j)=\left \langle n_{j}\right \rangle$ of interacting Fibonacci-Kitaev chain with $\mu = 4$, $ L = 610$, $\Delta = t$ and $U = 0.0, 1.6, 10.0$.
}
\end{figure}

The Fourier spectrum is thus obtained by taking fast Fourier transformation of the local occupation number $N(j)=\left \langle n_{j}\right \rangle$. The quasi-periodicity brought by the spatial varying potential can be most readily seen in Fourier spectrum \cite{diehl2003low}. For the Fourier spectrum of the $U = 0$ chain in Fig.~\ref{fig:Fourier}, the intensity is symmetric about $k = \pi$ and exhibits a fractal structure originated from the Fibonacci sequence. The fractal structure in the Fourier spectrum is destroyed by the interaction and evolved to the CDW structure. Increasing the interaction $U$, the prominent peaks around $k = \pi$ will get closer and have stronger intensity, and subsequently there exists only a single peak in Fourier spectrum at $k = \pi$ corresponding to an occupation number distribution of CDW phase.

\begin{figure}
\includegraphics{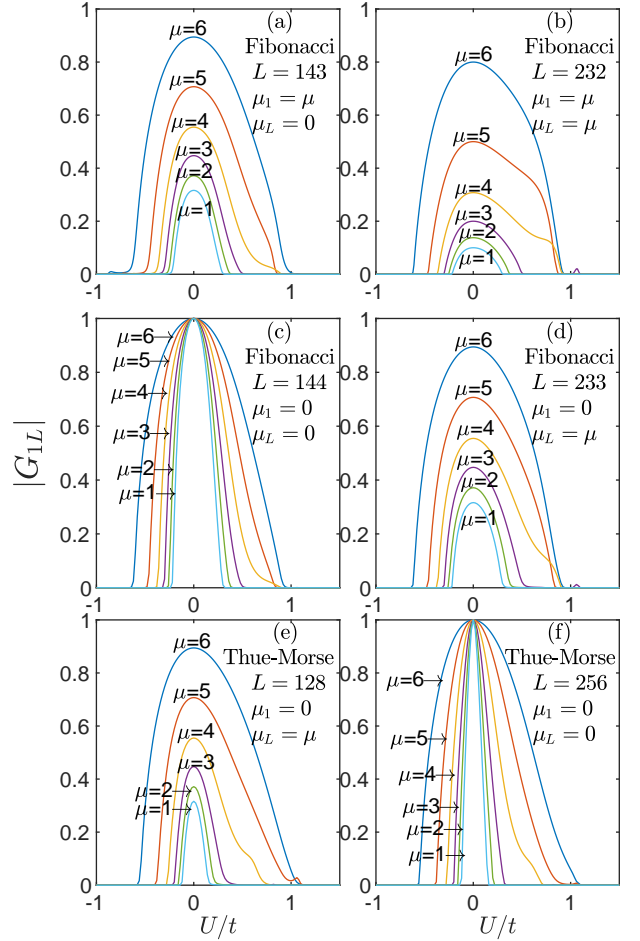}
\caption{\label{fig:EdgeTypeSixPic} Edge correlation function $G_{ij}$ of ground state of the interacting Kitaev chain with different types of chemical potentials. Whereas $\Delta = t$, $\mu_A =\mu$, $\mu_B = 0$. (a) Fibonacci lattice, $ L = 143$, $\mu_1 = \mu$, $\mu_L = 0$. (b) Fibonacci lattice, $ L = 232$, $\mu_1 = \mu_L = \mu$. (c) Fibonacci lattice, $ L = 144$, $\mu_1 = \mu_L = \mu$. (d) Fibonacci lattice, $ L = 233$, $\mu_1 = 0$, $\mu_L = \mu$. (d) Thue-Morse lattice, $ L = 233$, $\mu_1 = 0$, $\mu_L = \mu$. (d) Thue-Morse lattice, $ L = 233$, $\mu_1 = 0$, $\mu_L = \mu$.}
\end{figure}

The edge correlation function is sensitive to the chemical potential at the edge rather than the system size. In order to examine it, we use different length of Fibonacci sequence (up to $J_{13}$) and conclude that the chains with $\mu = 0$ at both boundary have edge correlation function $G_{1L} \approx 1$ when the interaction is small, which is distinct from the other three sets shown in Fig.~\ref{fig:EdgeTypeSixPic}. This is also valid to the chain with quasi-periodic chemical potential following other sequence such as the Thue-Morse sequence as shown in Fig.~\ref{fig:EdgeTypeSixPic}. For the Thue-Morse lattice, we use recursion formula $S_{n+1}=\{S_{n}, S_{n}^{-1}\}$, $n \geq 1$, $S_{0}=\{A, B\}$, so we have $S_{1}=\{S_{0}$, $S_{0}^{-1}\}=\{A, B, B, A\}$, $S_{2}=\{S_{1}$, $S_{1}^{-1}\}=\{A, B, B, A, B, A, A, B\}$, \dots. The total number of symbols is $S_{n}$ is $G_{n} = 2G_{n-1} = 2\times2^{n}$.
It is found that, although there are some differences for the four kinds of chemical potential at the edge respectively following the Fibonacci sequence and the Thue-Morse sequence, the latter leads to equivalent results and the exotic topological trivial phase does survive all of them.

Generally speaking, all the three potentials we are studying exhibit almost same behaviors with negligible distinctions. It means the configuration of potential and disorder does almost not matter. Merely the competition between the interaction and the chemical potential therefore can not explain the emergence of the symmetry breaking which is different from all other known situations. It deserves to imply adjacent sites with zero potential play the essential role. Namely, the occupation number on adjacent sites with zero potential in the chains suggests that the pairing mechanism of Fracton probably leads to the energy increase of the low-lying states with $Z_{2}^{f} = P = +1$. Hence, the $Z_{2}^{f}$ symmetry of the ground states is broken and the TSC phase turns into the trivial phase with four non-degenerate low-lying states. Fracton is a novel topological elementary excitation in fractal structures with dimension-limited mobility. Here in this work, we do not consider to calculate the dc conductivity on the lattice, so we are currently not able to determine whether the topological trivial phase is explicitly the Fracton phase. More investigations are reserved for the future dynamic researches.

\section{Conclusion}

In summary, we have studied the interacting Kitaev chains with periodic and quasi-periodic chemical potential by using the variational matrix product state (VMPS) method. In our innovative way to introduce the non-uniform chemical potential, we calculated the edge correlation fucntion $G_{1L}$ and found an newly-emergent topological trivial phase. Notably, symmetry is spontaneously broken in the topological superconducting (TSC) phase which is then split into two branches. This appealing phenomenon can be found in all the cases of spatially varying potentials. The two lowest states in each parity sector and occupation number are also calculated. The adjacent sites with zero chemical potential might be the key ingredient for the emergent phase.

\section*{Acknowledgment}

The authors gratefully acknowledge support from the National Natural Science Foundation of China (Grant Nos.~91833305, 11974118), Key Research and
Development Project of Guangdong Province (Grant No.~2020B0303300001), Guangdong-Hong Kong-Macao Joint Laboratory of Optoelectronic and Magnetic Functional Materials program (No.~2019B121205002) and Fundamental Research Funds for the Central Universities (Grant No.~2019ZD51).

\bibliography{apssamp}

\end{document}